# FROM SITE RESPONSE TO SITE-CITY INTERACTION: A CASE STUDY IN THE TOKYO AREA

P.-Y. Bard[1], K. Nakano[2], E. Ito[3], J. Sun[3], Z. Wang[3] & H. Kawase[3]

[1] ISTERRE, Grenoble, France, pierre-yves.bard@univ-grenoble-alpes.fr
[2] HAZAMA ANDO CORPORATION, Tsukuba, Japan
[3] Disaster Prevention Research Institute, Kyoto, Japan

**Abstract**: *Considering the purpose of the session relating early engineering developments in site response and soil-structure interaction, this paper focuses on the development of studies regarding site-city interaction following the striking site response observations obtained in Mexico City during the 1985 Guerrero-Michoacan event, The first part presents an overview of the investigations on multiple structure-soil-structure interaction, starting with Mexico-city like environments with dense urbanization on soft soils, which later evolved with the concept of metamaterials. Up to now, such investigations have been largely relying on numerical simulations in 2D and 3D media, coupling soft surface soil layers and simplified building models, including also some theoretical developments using various mechanical concepts. They also relied on a number of laboratory experiments on reduced-scale mock-ups with diverse vibratory sources (shaking table, acoustic devices). The latest studies coupled full-scale experiments on mechanical analogs such as forests or wind turbine farms involving sets of resonators with similar frequencies, and numerical simulation to investigate their impact on the propagation of surface (Rayleigh) waves. Almost all such studies converge in predicting lower ground motion amplitude for sites located within the "urbanized" area, but none of them can be considered a "ground-truth" proof for a real earthquake in a real city. The second part thus takes advantage of the long duration of strong motion observations in the Kanto area thanks to the KiK-net, K-NET and JMA (Shin-dokei) networks, to investigate the possible changes in site response with time. The first results obtained with the event-specific site terms derived from Generalized Inversion Techniques (Nakano et al., 2015) indicate a systematic reduction of the low frequency (0.2 – 1 Hz) site amplification, in the central-south Tokyo area. As this frequency band corresponds both to the site frequency (very thick deposits) and to the high-rise buildings, the discussion focuses on the possible relation with the extensive construction in some areas of downtown Tokyo over the last 2 decades.*

## 1  Introduction

The population growth and rural exodus has resulted in the urban densification of many large cities, established on soft soil areas along rivers or sea-shore, with more and more high-rise buildings. Such a concentration of population and various types of human activities in urban centers raises the attention on the associated seismic risk, the estimation of which is most often performed in a simple sequential way: the seismic hazard depending on source characteristics, crustal propagation properties and site conditions is characterized in terms of free-field ground motion, which is then considered as an input for estimating the behavior of man-made constructions. While the latter step may account for soil-structure interaction (SSI) effects for large-size buildings located on soft soils, not only the presence of nearby constructions and the possible structure-soil-structure interaction (SSSI), but also the possible feedback of building clusters on the "free-field" motion within urbanized areas, are most often neglected.

The structural engineering community has very early launched many investigations to understand and model the effects of soil-structure interaction (SSI) on the structural response. Comparatively much less attention has been paid to the possible feedback of multiple SSI on the "free-field" ground motion, i.e., on the site response. The first studies of the mid-eighties (Yamabe et al., 1986; Yamabe and Kanai, 1988) focused on an apparent increase of ground motion attenuation with distance in the (urbanized) Tokyo-Yokohama area compared with the rest of Japan. From another viewpoint, the very specific waveforms exhibiting long duration beating phenomena that were recorded in the Mexico city lake bed area during the 1985 Guerrero-Michoacan earthquake, triggered a series of investigations aiming at quantifying the scattered wavefield radiated away from the foundations of vibrating high-rise buildings. The first part presents an overview of the various approaches used in that aim (numerical simulation, laboratory experiments at reduced scale using centrifuge, shaking tables or acoustic devices, or at full scale on analogues such as forests or wind farms) and of the corresponding results, indicating the plausibility of measurable, significant effects on ground motion, with an overall decrease of the average ground motion in specific frequency bands together with an increased spatial variability. However, such effects could never be unambiguously proved using seismological recordings from real earthquakes in real cities. The multiplication of strong motion instruments installed and maintained for a long time within urban environments allows now to start tackling this issue: the second part advantage of the long duration of strong motion observations in the Kanto area thanks to the KiK-net, K-NET and JMA (Shin-dokei) networks, to investigate the possible changes in site response with time. The first results obtained with the event-specific site terms derived from Generalized Inversion Techniques (Nakano et al., 2015) indicate a systematic reduction of the low frequency (0.2 – 1 Hz) site amplification, in the central-south Tokyo area. As this frequency band corresponds both to the site frequency (very thick deposits) and to the high-rise buildings, the following discussion section addresses the major urban changes occurred in some areas of downtown Tokyo over the last 2 decades, and the possible relation with the observed reduction in low-frequency site response, while the concluding comments propose a list of further investigations to support or weaken such a "site-city interaction" interpretation.

## 2    From site effects in Mexico City to investigations on site-city interaction (SCI)

The 1985 Guerrero-Michoacan earthquake and its long-distance destructive effects in the lake-bed zone of Mexico City produced very rich and unexpected observations, which in turn triggered a lot of investigations. On one side, the large, very broad band, amplification factors derived from strong motion recordings modified the perception of "beneficial" effects of soil non-linearity, leading to significant evolutions in building code spectral shapes and amplitudes for various site categories, On the other hand, the long duration and beating shape of some waveforms obtained in the lake-bed zone (Singh & Ordaz, 1993), could not be satisfactory explained on the basis of 1D or 2D site-effects only (Chavez-Garcıa & Bard, 1994). As beatings are often observed in structural recordings because of the existence of different modes with close natural frequencies, some authors put forward the possible feedback from nearby buildings to explain such observations. This "single-scattering" topic had already been addressed previously, but it was further investigated and lead over the years into the exploration of more global interactions between building clusters and underground structure, which was called "site-city" interaction. The present section briefly presents an overview of these investigations and of their outcomes.

### 2.1    Radiated wavefield from a single building

The effects of soil-structure interaction on the structural response have been recognized for a long time by the civil engineering community (e.g., Jennings and Bielak, 1973), including the "radiative damping" which increases the apparent damping through wave radiation back into the foundation soil. However, not much attention was paid on the effects on "free-field" ground motion even near the building base, as it was considered that the energy radiated away from the building was decaying very rapidly with distance.

Some observations were however obtained for some single buildings forced into vibrations by some veru specific excitations, indicating that the radiated could be detected even at large distances. The first observation was reported by Jennings (1970) during forced vibration tests with actuators placed at the roof of the Millikan library building on the Caltech campus: seismographs located a few kilometers away recorded harmonic ground motion corresponding to the building period. The same experiment was repeated several times, and during one of them, similar observations could be performed at many stations of the southern California broad-band network, up to a 100 km away (Favela, 2004). The second one was associated with the shock waves created by the reentry into the atmosphere of the Columbia space shuttle on its way back to Edwards Air Force base in California: Kanamori et al. (1991) identified specific arrivals at some broad-band stations in the Los Angeles area as due to "a seismic P wave excited by the motion of high-rise buildings in downtown Los Angeles, which were hit by the shock wave" and considered that "the proximity of the natural period of the high-rise buildings to that of the Los Angeles basin enabled efficient energy transfer from shock wave to seismic wave". Similar observations were also obtained on September 11, 2001 terrorist attacks in New-York state: the two plane impacts against the World Trade Center were recorded in several seismological stations at distances of several tens of kilometers (Kim et al., 2001).

Several specific experiments were also performed in order to detect such radiated waves, and confirmed their potential significance. This was done for instance by Gueguen et al. (2020) with pull-out tests on a reduced scale (1/3) 5-story frame structure built at the Volvi test site near Thessaloniki in Greece (waves could be detected up to 50 m from the 5m high model structure) , or by Cardenas et al. (2000) and Chavez-Garcia and Cardenas-Soto (2002) investigating the ground motion around instrumented Jalapa building in Mexico City. On the contrary, Castellaro and Mulargia (2010) reported three cases in Italy (three famous towers in Pisa, Venice and Bologna) where radiated waves could hardly be detected at distances larger than 12 m. Nevertheless, using some innovative processing combining deconvolution and polarization analysis, Skłodowska et al. (2021) could textract radiated waves from earthquake recordings in the Matera (Italy) area, and conclude that the energy transmitted from vibrating building to its surroundings is actually significant, leading to a decrease of ground motion because of out-of-phase motion: This might explain why such radiated waves are uneasy to detect when looking only at signal amplitude.

On this basis, Gueguen et al. (2002) developed a simple SSI model based on impedance functions and Green's functions at the surface of a layered half-space, and calibrated it on the data from the pull-out experiments at Volvi test site. They used this model to estimate the cumulative contributions, at a given location within the Colonia Roma area in Mexico City, of all secondary waves radiated from 180 high-rise buildings (i.e., exceeding 7 stories) located within a distance of 500 m. This simplified computational approach allowed to reproduce the long duration and beating of the observed ground motion in Mexico City. It was however too simple as it neglected multiple structure-soil-structure interaction effects: in such a simple "single-scattering" approach, the total wavefield radiated back into the soil from the buildings is considered to be the superposition of the effects of each individual building, which are (wrongly) supposed not to interact with each other. Neighbouring buildings can actually exchange non-negligible amounts of energy with each other, and constitute, through the soil, a fully coupled dynamic system that should be treated as a whole.

## 2.2   Multiple structure-soil-structure interaction

Analyzing the effects of structure-soil-structure interaction within building clusters located on vertically and sometimes laterally heterogeneous soils is not straightforward. Historically, it was first approached with numerical simulation, which became increasingly complex with the exponential growth of computing power. For some specific building layouts (e.g., with 2D or 3D periodicity), some theoretical models based on homogenization theory could be developed, providing a physical insight on the phenomena at play in such interactions and on the relevant mechanical parameters. Over the last two decades, these numerical results and theoretical models could be compared with the observations obtained in specifically designed laboratory experiments on small-scale analog models (centrifuge, shaking table, acoustic experiments), and more recently (i.e., the last decade) in some "full-scale, in-situ" measurements using some simple analogs for buildings and cities (i.e., trees and forests, or windfarms).

*Numerical simulations*

The first issue to be formulated was the interaction between two neighboring buildings (Luco & Contesse, 1973). It was then extended to multiple interactions with building groups by Wong & Trifunac (1975), This issue received a special attention in nuclear engineering because of the frequent co-location on the same site of several nearby nuclear reactors (see Lou et al., 2011 for a review). However, not much attention was paid to the ground motion for sites located within the structural clusters, until Wirgin & Bard (1996) who suggested, on the basis of rather simplistic 2D models, that such collective interactions among buildings, could provide a possible explanation for the atypical features observed in Mexico City recordings during the 1985 Guerrero-Michoacan earthquake.

It then gave rise to an increasing number of numerical simulations based on various computational methods (finite elements, boundary elements, spectral elements, Green functions, etc.), which were applied to small groups of nearby buildings or to simplified city models (2-D or 3-D, with periodic or quasi random building layouts). Clouteau & Aubry (2001) presented pioneering 3D computations for the Roma-Norte area in Mexico City, which were the first to indicate an overall average decrease of ground motion within the built area, with however an increased spatial variability due to the multiple constructive and destructive interferences between all the waves (multiply) radiated from each single building (Ishizawa et al., 2003). Their results were generally confirmed by a number of later publications considering completely idealized urban environments and underground structures (Tsogka & Wirgin, 2003; Groby et al., 2005;; Ghergu & Ionescu, 2009; Padron et al., 2009; Iqbal et al., 2012; Iqbal, 2014; Sahar et al., 2015; Kumar & Narayan, 2018, 2019; Tian et al., 2020; Basnet et al., 2023), or idealized urban settings on real underground structures (Kham et al., 2006; Semblat et al., 2008; Taborda and Bielak, 2011; Isbiliroglu et al., 2015; Kato, 2020 and Kato and Wang, 2021, 2022), or realistic cities (Uenishi et al., 2010 for the Friuli, Italy, area; Guidotti et al., 2012 for the central Business District in Christchurch, New-Zealand; Lu et al., 2018 for the Tsinghua University campus in Beijing, China; Varone et al., 2021 for the Fosso di Vallerano valley in Rome, Italy).

Almost all these simulations support the idea that cities can interact with the underground site structure as a whole, especially when the buildings and site frequencies correspond. Apart from a slight shift of the apparent site fundamental frequency towards lower values, leading to some amplification in a narrow frequency band just below the site natural frequency (added mass effect), the average effect is a decrease of ground motion (and building response) within the urbanized area, together with an increased spatial variability, possibly leading to some amplifications at some places, the location of which is however highly dependent on the incoming signal characteristics (waveform, incidence and azimuth for plane waves, source location for point or extended earthquake-like sources). These global interaction effects are the largest for homogeneous and periodic buildings having a frequency tuned to the site natural frequency, and decrease both with increasing heterogeneity in the building spatial and geometrical characteristics (which is usually the case for real cities) and with increasing differences between site and building frequencies.

*Analytical models*

Alternative, analytical approaches were developed in parallel by Boutin & Roussillon (2004, 2006) to investigate the effective behavior of idealized, periodic city arrangements on the basis of the homogenization approach. When the representative surface element of the city (constituted by one or few buildings) is small compared to the seismic wavelength, the city can be described as a surface impedance that modifies the usual free-field condition (Schwan & Boutin, 2013; Boutin et al.,2015) and the resulting models yield ground and building motion trends which are very similar to those derived from much more complex numerical simulations.

*Experiments*

The first dedicated experiments delt with small-scale specimens consisting of a limited number (2 to 5) of reduced-scale closed-spaced oscillators mimicking buildings excited by a shaking table (Kitada et al. 1999) or by small ball shots inside a centrifuge (Chazelas et al. 2003). Both concluded that (multiple) structure- soil-structure interactions have not only observable, but also noticeable effects on the dynamic response of the structures and ground motion. More recently, Schwan et al. (2016) carried out several shaking table test with numerous oscillators, which allowed both to validate the homogenization-based surface impedance models, and to identify the key parameters controlling the main characteristics of site-city interaction in the low frequency range (wavelength much larger than building spacing, which were one again found to be related to the stiffness and mass properties of both buildings and underlying soils, the (reduction) effects being the largest for coinciding frequencies.

The latest large-scale experiments dealt with almost full-scale, in -situ analogues: forests with the "metaforet" project (Colombi et al., 2016; Roux et al., 2018) and windfarms (Mohammed, 2024; Pilz et al., 2024). Both experiments were designed to explore the fact, found in lab-scale experiments involving metamatrial plate and rods (Colombi et al., 2015, 2016) that resonators such as trees or windturbines result in forests or windfarms acting as locally resonant metamaterials, with important bandgaps for the propagation of Rayleigh surface waves. The "metaforêt" experiment did demonstrate that Rayleigh waves, propagating in soft sedimentary soils experience strong attenuation, when interacting with a forest, over two separate large frequency bands, which are related to the interaction between longitudinal resonances of trees with the vertical component of the Rayleigh wave. A similar experiment was carried out recently in Germany, consisting of a dense grid of 400 geophones installed inside and outside a windfarm (Pilz et al., 2024; Mohammed, 2024). Analyzing the seismic wavefield propagating through the wind farm showed that surface waves are strongly damped in several frequency bands in relation to flexural and compressional resonance modes of the wind turbines. The associated reduction effects are significant, but they involve only Rayleigh waves, which are only one component of actual earthquake wavefields which also consist of body and Love waves.

All these results seem however consistent at least to justify the search for noticeable effects on seismic wavefield and ground motion characteristics in densely urbanized areas. Because of the difficulty to identify with enough details all the components of a seismic wavefield even outside of any urban environment (e.g., Imtiaz et al., 2020), it is easier to look for site response changes in changing urban environments: this is the purpose of the next section, considering the Tokyo area which has been continuously monitored for several decades, and where a few central wards experienced considerable re-development in recent decades.

## 3   The Tokyo case Study

Despite the consistent evidence of the reality of multiple SSI from numerical simulation, small scale laboratory experiments or full-scale analogs (forests, wind farms), the effectiveness of global site-city interaction could never be proved explicitly in actual cities with actual seismological recordings recorded during real earthquakes. The purpose of the present section is to investigate whether some changes can be detected in one of the most densely urbanized, most active and instrumented areas of the world, the Tokyo area.

### 3.1   Data set

The whole Japan has benefitted for the KiK-net and K-NET instrumentation program launched just after the Kobe (1995) event, which offers a continuous set of recordings for the last 2-3 decades since 1996. This data set, augmented by the JMA (Shin-dokei) networks has been used by several authors to perform generalized inversions in order to extract the source, path and site terms for each event and station. In the present study, we use the latest site terms obtained with the methodology described in Nakano et al. (2015), on the subset of data focusing on the Kanto area. These site terms are estimated with respect to the "seismological bedrock" with a S-wave velocity of 3500 m/s (corresponding to the bedrock at the selected reference station YMGH01). We consider all the K-NET, KiK-net and JMA sites located in the four prefectures comprising and surrounding the central business district of Tokyo and its immediate neighborhood, including the renewal areas where a large number of high-rise buildings (with height exceeding 150 m) has been constructed at the dawn of the XXI$^{st}$ century. This dataset corresponds to a total of 148 sites as shown in Figure 1, consisting of 128 K-NET and KiK-net sites (24 in the Kanagawa prefecture, 26 in Saitama, 42 in Chiba and 36 in Tokyo), and 20 JMA sites located mainly in the Kanto plain. Most of the K-NET and KiK-net instruments were installed in the late 90's, while the JMA stations cover very diverse periods, starting sometimes earlier, most often later, and for limited durations only.

In their approach, Nakano et al. (2015) used either S-wave portions with a limited duration (at most 40 s for the largest events) or S-wave and surface-wave parts corresponding to much longer signal duration (up to 655 s). Considering the location of many stations within the Kanto plain with energetic, long-period surface waves generated on the edges of the Kanto plain (Kinoshita et al., 1992), we used the latter results which exhibit larger amplification especially in the low-frequency range, as displayed in Figure 2 for a typical example station located in downtown Tokyo. This choice is also supported by the fact that several recent studies (Colombi et al., 2016; Roux et al., 2018; Pilz et al., 2024; Shoaib et al., 2024) emphasized the bandpass reduction role of sets of similar resonators (trees, wind turbines) on the propagation of Rayleigh waves, and from the original findings by Yamabe & Kanai (1988) reporting a larger attenuation within the Tokyo urban area.

The core of the present study consists first in investigating whether the so-extracted site terms (which correspond to the frequency range from 0.2 to 20 Hz) exhibit a statistically meaningful change with time, and second, in case of changes, in discussing whether these changes might be related with the local urban development occurred over the considered time period. In order to have reliable results, we decided to keep only sites with recordings over a time period longer than 10 years, which eliminated 42 stations (including all the JMA stations). For the remaining 106 sites, the average number of available recordings is 131, with a large standard deviation (± 60).

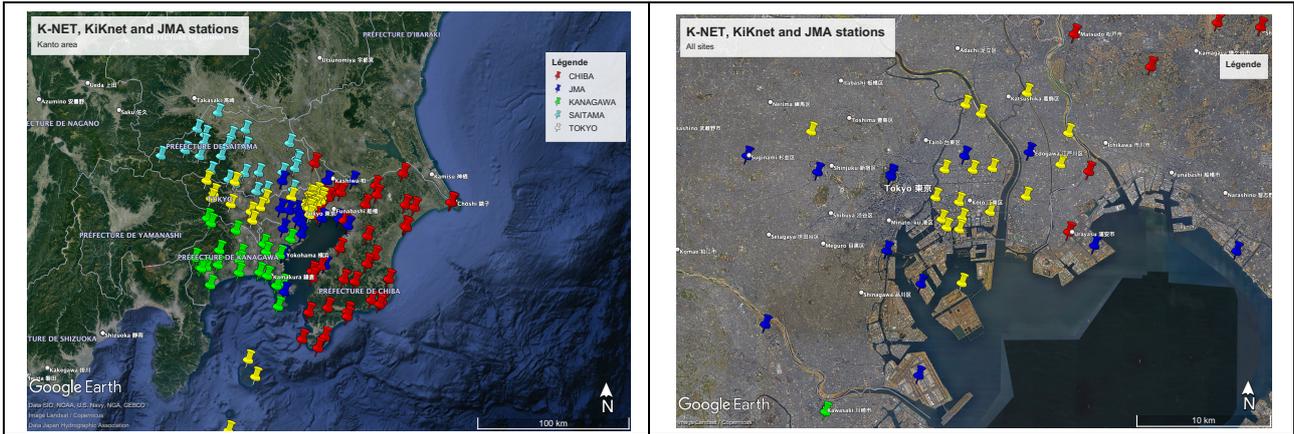

*Figure 1: Map of considered stations in the Tokyo area. On the left are displayed all the K-NET and KiK-net stations of the Tokyo (yellow), Chiba (red), Saitama (cyan) and Kanagawa (green) prefectures, together with the JMA stations (dark blue). The right panel focuses on the Tokyo downtown area.*

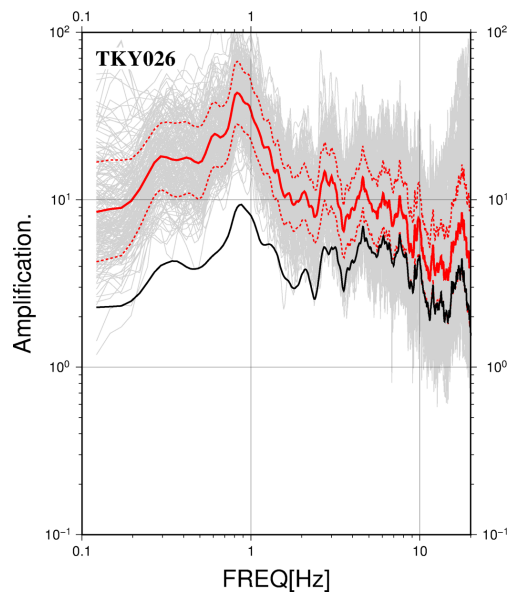

*Figure 2: Example site terms for the TKY026 (K-net) station: grey curves correspond to individual site terms, red curves to the geometrical average (solid line for mean, dotted lines for mean ± 1 standard-deviation), obtained when considering the long S-wave + surface wave part. For comparison purpose, the average site term obtained on the same recordings when considering only the S-wave part is also shown in black.*

### 3.2 Looking for time changes: Data processing

The potential site response changes have first been investigated in a semi-qualitative, semi-quantitative way by comparing the average site response terms over different time periods. Figure 3a shows example results for the same station TKY026, comparing the (geometrical) average site term computed over all available recordings (198 for this station, black solid line) with the similar average computed over four successive time periods: blue for the period until 2002 (14 recordings), green for the 2003-2007 period (39 recordings), magenta for the 2008-2012 period (93 recordings), and red for the period since 2013 (52 recordings). The most remarkable feature on this result is the stability of the site term for frequencies between 2 and 10 Hz, and its steady decrease with time for frequencies below 1 Hz. In order to check the statistical meaning of such a decrease, a Student test was performed for each pair of time-period average, to estimate the probability they are the same, considering the number of recordings and the associated standard-deviation for each frequency and time period. The color curves on the bottom plot of Figure 3a show, for each time period (and the same color code as the top plots), the minimum probability value that it is similar to the average in another time period. To provide an idea of the event-to-event variability of the site term, the black curve in the same plot shows the log10 standard-d                         e minimum value is around 0.15 for
intermediate frequency range                       Hz and above 8 Hz, and exceeds 0.25
for very low and high frequenc                     ectively). It turns out that the probability
of similar average over differe                    s between 1.5 and 10 Hz, and very low
(below 0.01) for frequencies b                     lso decreasing beyond 10 Hz, but this
high-frequency behavior may                        actors which cannot be controlled and
monitored with the presently a

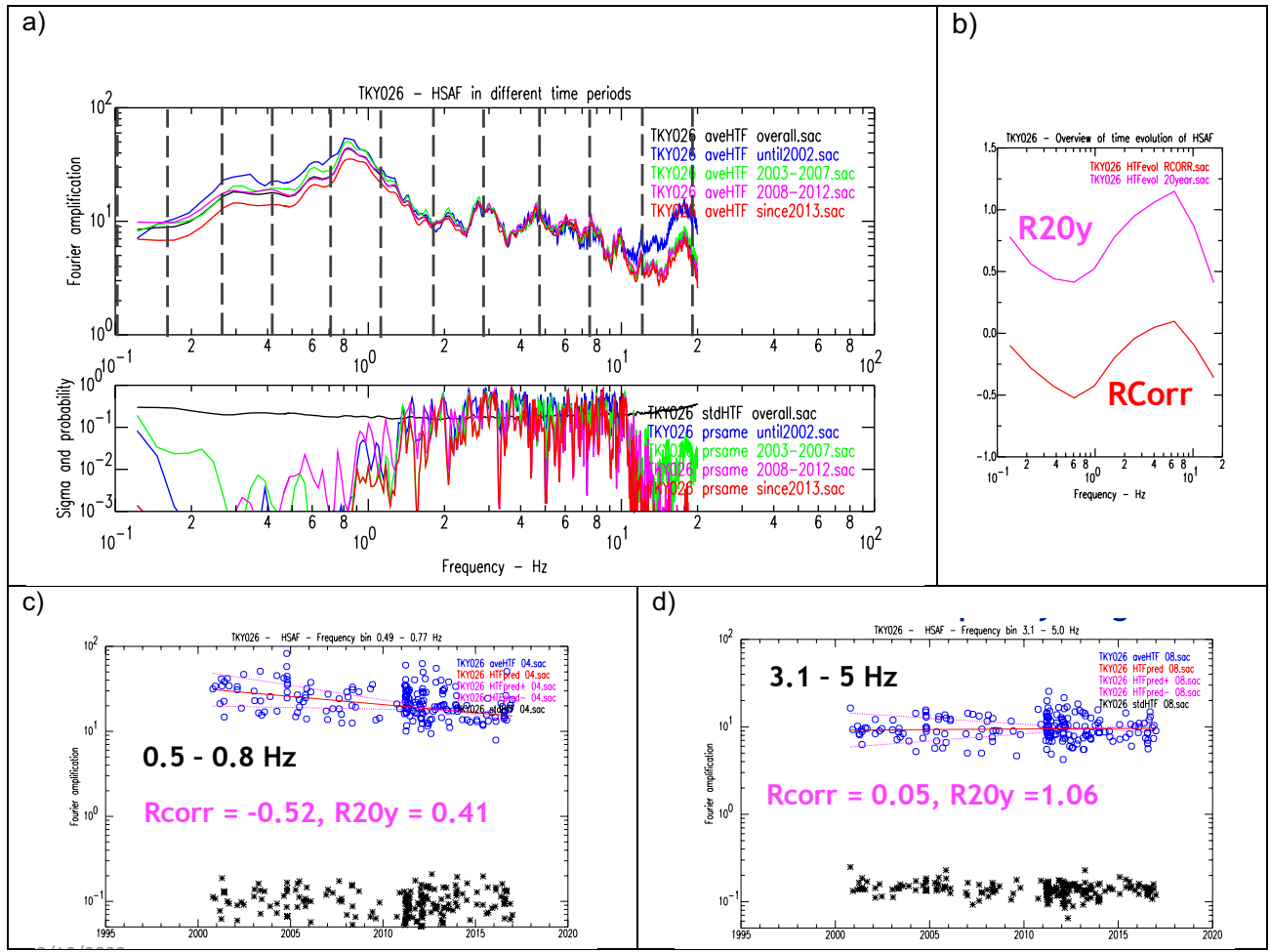

*Figure 3: Example processing for station TKY026. On top left (a) are displayed the site response average over four different time periods (blue/green/magenta / red from older to most recent), together with the overall average (black solid line); the vertical dashed lines indicate the 11 frequency ranges over which site term are averaged for further analysis; The plot below displays the log10 standard deviation for the overall average, and the probability that each of the 4 individual average are the same as the overall average (with the same color code). The two bottom plots (c and d) show the evolution with time (year in abscissa) of the site terms for each event (blue open circles) for two specific frequency bands (0.5-0.8 Hz on the left, and 3.1 – 5 Hz on the right), and the corresponding log-linear trend (magenta, solid line = average, dashed lines = average ± one standard-deviation); the bottom black symbols represent the standard deviation for the site term averaged over the considered frequency band. The top right plot displays the evolution with frequency of the correlation coefficient Rcorr and of the 20 year reduction ratio R20y corresponding to the log-linear regressions displayed in plots c and d. See the main text for further details.*

Given the existence of statistically meaningful changes over time, a further, more quantitative analysis was performed by considering 11 frequency bins with an equal geometrical spacing from 0.12 to 20 Hz: these frequency bins are indicated by the vertical dashed black lines in Figure 3a, and correspond to central frequencies of 0.16, 0.25, 0.40, 0.64, 1.0, 1.6, 2.5, 4.0, 6.4, 10.0, and 16 Hz. For each recording i of the 198 events obtained at this site, a geometrical average site term $ST_{ij}$ was computed for each bin j, together with the corresponding standard deviation, and its evolution with time was analyzed as shown in Figure 3b and c for two example frequency bins in the low frequency (bin 4, 0.49-0.77 Hz, Figure 3b) and intermediate frequency ranges (bin 8, 3.1-5 Hz, Figure 3c). In addition, a log-linear regression was performed to quantify the time dependency of $\log_{10}(ST_{ij})$ through two quantities: the correlation coefficient Rcorr providing an indication on the reliability of the (positive or negative) variation with time, and the value R20y providing the reduction factor over a 20 year period (derived from the slope of the regression). The two example plots are representative of the two different behaviors, with a rather strong (negative) correlation at low frequency (Rcorr = -0.52), and an associated 20 year reduction of 2.5 (R20y = 0.4), and no correlation (Rcorr = 0.05) and no 20 year trend (R20y =1.06).The top right plot in Figure 3d summarizes the results for the 11 frequency bins, showing the evolution with the bin central frequency of the Rcorr (red line) and the R20y (magenta line) values. The Rcorr values are below -0.4 only for bins 3 to 5, i.e., for frequencies from 0.3 to 0.8 Hz, and correspond to very weak correlations (between -0.2 and +0.2) for bins 6 to 10 (i.e., for frequencies between 1.4 and 12 Hz), with R20y values around 1 (no change with time).

### 3.3  Results: Identification of stations with changing site response

*Statistical relationship between Rcorr and R20y*

Similar analysis were performed for the 106 stations offering more than 10 years of recording. Before analyzing the sites with actual changes, it is useful to have a look at the statistical relationships between the values of the Rcorr correlation and the 20 year reduction factor R20y. which may be derived from the 106 pairs of (Rcorr, R20y) values and for the 11 frequency bins. As expected from their calculation procedure, these pairs are highly correlated for each frequency bin. Figure 4 shows such relationships for two representative frequency bins, one for the low frequency range (bin 3, 0.3-0.5 Hz) and one for the intermediate frequency range (3 – 5 Hz), which calls for several comments:

- There exist a significant number of stations exhibiting large negative Rcorr values (i.e., below -0.4), essentially at low frequency, while very few stations exhibit large positive Rcorr values (i.e., larger than 0.4), whatever the frequency bin.
- Similarly, there exist several stations for which the low frequency R20y value is below 0.33, corresponding to an amplification reduced by at last 3 over a 20 year period.
- The slope of the (Rcorr, R20y) correlation plot is changing from one frequency bin to another, with slightly larger slopes at low frequency: a Rcorr value of -0.4 (respectively -0.5) is associated to R20y values around 0.4 (resp. 0.35) at low frequency, and 0.5 (resp. 0.4) at high frequency

Rcorr and R20y are thus almost equivalent ways to characterize the evolution with time of the site response. Moreover, given the fact that most of the changes consist of a reduction, it was decided to use the inverse of R20y to characterize them: this factor is noted DF20 (=1/R20y) and is called the "20 year decrease factor".

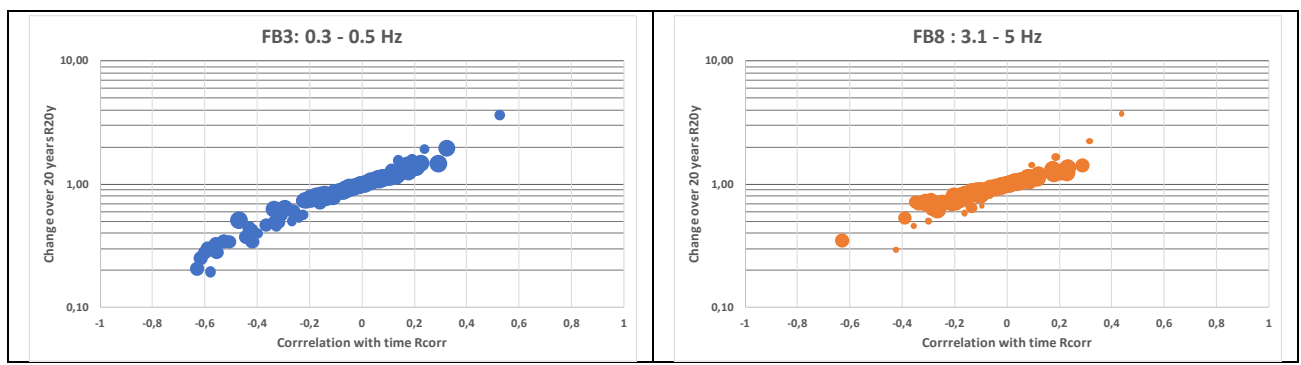

*Figure 4: Correlation between Rcorr and R20y values for two example frequency bins (bin3, 0.3-0.5 Hz on the left, and bin 8, 3.1-5 Hz on the right). The size of the symbols for each data point is proportional to the total duration of observations (between 10 and 20 years)*

*Stations with low-frequency changes*

We have first looked for stations exhibiting a "consistent, significant" low-frequency decrease, i.e. stations for which one low-frequency bin (below 1Hz) corresponds to a Rcorr value lower than -0.3 and a mean low-frequency Rcorr value smaller than -0.2, or almost equivalently, a mean DF20 value above 1.7 and at least one bin for which it is above 2. A total of 15 stations in the Kanto area fulfill these criteria, the corresponding DF20 and Rcorr values are listed in Table I for the four low frequency bins from 0.2 to 1 Hz, while all the other stations do not exhibit any statistically meaningful change (decrease or increase). This list can be split in four groups:

- 3 sites (TKY020, TKY018 and TKY016) exhibit a large reduction with an average reduction factor DF20 above 3
- 5 sites (TKY014, TKY015,, TKY027, TKY023 and TKY021) exhibit a moderately large reduction with an average low-frequency reduction between 2.5 and 3
- 4 sites (TKY022, TKY017, TKY026 and TKY028) exhibit a noticeable low-frequency reduction between 2.0 and 2.5
- Finally, 3 sites (TKY013, TKY024 and TKYH13) exhibit a slight reduction, possibly statistically meaningful, reduction between 1.6 and 2

*Table I: Summary information for the stations with remarkable low-frequency decrease.*

| Station | Log10(DF20)/ bin | | | | Mean | DF20 | Rcorr | | | | Mean |
|---|---|---|---|---|---|---|---|---|---|---|---|
| | Frequency (Hz) | | | | | | Frequency (Hz) | | | | |
| | 0.25 | 0.40 | 0.64 | 1.0 | | | 0.25 | 0.40 | 0.64 | 1.0 | |
| TKY020 | 0.510 | 0.684 | 0.592 | 0.355 | 0.535 | 3.428 | -0.601 | -0.629 | -0.655 | -0.507 | -0.598 |
| TKY018 | 0.577 | 0.597 | 0.536 | 0.367 | 0.519 | 3.304 | -0.682 | -0.616 | -0.623 | -0.557 | -0.620 |
| TKY016 | 0.489 | 0.551 | 0.484 | 0.479 | 0.501 | 3.169 | -0.534 | -0.554 | -0.585 | -0.617 | -0.573 |
| TKY014 | 0.556 | 0.520 | 0.456 | 0.336 | 0.467 | 2.931 | -0.543 | -0.588 | -0.585 | -0.462 | -0.545 |
| TKY015 | 0.542 | 0.551 | 0.458 | 0.285 | 0.459 | 2.879 | -0.551 | -0.599 | -0.536 | -0.425 | -0.528 |
| TKY027 | 0.554 | 0.465 | 0.409 | 0.303 | 0.433 | 2.708 | -0.509 | -0.508 | -0.494 | -0.399 | -0.478 |
| TKY023 | 0.487 | 0.483 | 0.474 | 0.282 | 0.431 | 2.700 | -0.535 | -0.558 | -0.561 | -0.403 | -0.514 |
| TKY021 | 0.463 | 0.465 | 0.401 | 0.315 | 0.411 | 2.577 | -0.503 | -0.528 | -0.512 | -0.477 | -0.505 |
| TKY022 | 0.264 | 0.466 | 0.478 | 0.362 | 0.392 | 2.467 | -0.236 | -0.42 | -0.499 | -0.472 | -0.407 |
| TKY017 | 0.381 | 0.426 | 0.329 | 0.180 | 0.329 | 2.132 | -0.413 | -0.443 | -0.437 | -0.281 | -0.394 |
| TKY026 | 0.250 | 0.356 | 0.383 | 0.283 | 0.318 | 2.080 | -0.28 | -0.43 | -0.523 | -0.428 | -0.415 |
| TKY028 | 0.313 | 0.374 | 0.322 | 0.240 | 0.312 | 2.052 | -0.357 | -0.42 | -0.429 | -0.375 | -0.395 |
| TKY013 | 0.278 | 0.331 | 0.289 | 0.120 | 0.254 | 1.796 | -0.322 | -0.366 | -0.395 | -0.202 | -0.321 |
| TKY024 | 0.338 | 0.310 | 0.231 | 0.119 | 0.250 | 1.777 | -0.346 | -0.321 | -0.246 | -0.136 | -0.262 |
| TKYH13 | 0.394 | 0.299 | 0.140 | 0.118 | 0.238 | 1.729 | -0.388 | -0.267 | -0.137 | -0.135 | -0.232 |

The location of these sites is shown in Figure 5a (except for the station TKYH13 which is located on the western border of the Kanto basin). All stations with at least significant reduction (i.e., above a factor of 2) are actually in the central area, with the largest reductions located to the south-East of the main railway station and in-between the Sumida and Arakawa rivers.

*Stations with high-frequency changes*

Similarly, we have looked for stations with consistent, statistically meaningful high-frequency decrease, with the same criteria as for the low-frequency case, i.e. stations for which at least one high-frequency bin (between 2 and 12 Hz) corresponds to a Rcorr value lower than -0.3 and a mean low-frequency Rcorr value smaller than -0.2, or almost equivalently, a mean DF20 value above 1.7 and at least one bin for which it is above 2. Only 4 stations in the Kanto area fulfill these criteria, the corresponding DF20 and Rcorr values are listed in Table II for the four high frequency bins from 2 to 12 Hz.

No station exhibits an average high-frequency reduction factor larger than 3, only one station (TKY016) exhibits a moderately large reduction (between 2.5 and 3), and the latter 3 (TKY020, TKY015 and TKY023) correspond to rather poor average trends with average reduction factor between 1.7 and 2. Within these 3 stations, only TKY020 displays a consistent decrease over the whole 2-10 Hz frequency range, while for the latter two, only the last bin around 10 Hz is characterized by a statistically meaningful reduction. All the other stations of the whole Kanto area do not exhibit any statistically meaningful change (decrease or increase) in this frequency range. A number of stations do exhibit a marked change (increase or decrease) in the last frequency bin (centred around 16 Hz), but such very high-frequency changes may be due to modifications in the sensor installation or in the immediate surroundings (see Hollender et al., 2020): as they are beyond our knowledge, we preferred not to take this bin into account here).

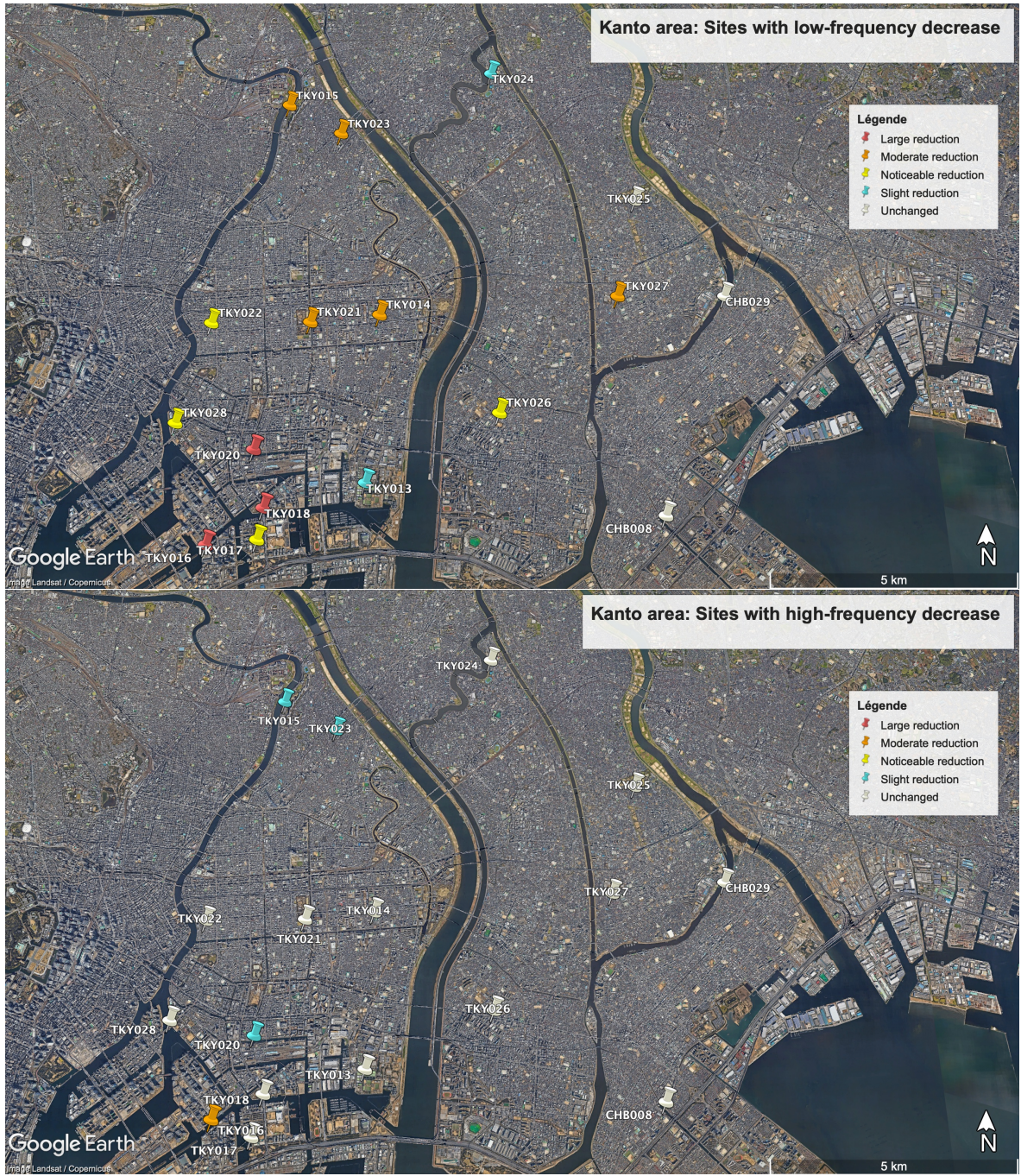

*Figure 5: Map showing the location of stations with statistically meaningful low frequency decrease (a, top) and high frequency decrease (b, bottom)*

*Table II: Summary information for the stations with remarkable high-frequency changes (only decrease).*

| Station | Log10(DF20)/ bin | | | | Mean | DF20 | Rcorr | | | | Mean |
|---|---|---|---|---|---|---|---|---|---|---|---|
| | Central frequency (Hz) | | | | | | Central frequency (Hz) | | | | |
| | 2.5 | 4. | 6.4 | 10 | | | 2.5 | 4. | 6.4 | 10 | |
| TKY016 | 0.401 | 0.456 | 0.502 | 0.526 | 0.471 | 2.959 | -0.55 | -0.63 | -0.602 | -0.639 | -0.605 |
| TKY020 | 0.276 | 0.144 | 0.208 | 0.519 | 0.287 | 1.935 | -0.504 | -0.347 | -0.454 | -0.6 | -0.476 |
| TKY015 | 0.199 | 0.089 | 0.086 | 0.572 | 0.237 | 1.725 | -0.362 | -0.182 | -0.169 | -0.601 | -0.329 |
| TKY023 | 0.135 | 0.148 | 0.149 | 0.485 | 0.229 | 1.695 | -0.194 | -0.252 | -0.217 | -0.587 | -0.313 |

The location of these four sites is shown on Figure 5b on a map similar to the one shown on Figure 5a, together again with all the stations of the same area which do not exhibit any change (white pins). The areas concerned by such potentially significant, changes in the intermediate to high frequency domain (TKY016 and TKY020) are both located south-East of the main railway station, but there are other sites in the same area that do not experience any change.

## 4    Discussion: could it be explained by site-city interaction ?

The main outcome of this analysis is that there exist a group of stations, all located in the same area, for which the site response has consistently decreased in the low-frequency range (0.2 – 1 Hz) over the last two decades. Before considering other potential explanations of such a decrease, we investigate here whether it could be due to some kind of site-city interaction in relation to the urban changes occurred in the Tokyo area over the same period. The next subsections therefore present some hints on the recent urban changes in the Tokyo downtown area and qualitatively discuss their potential impacts on site response, before considering some other possible explanations

### 4.1    Recent urban changes in the central Tokyo area

The central Tokyo area has experienced recently a major urban revitalization which led to the construction of a large number of high-rise buildings within the framework of the urban Renaissance program (Fujii et al., 2006; Languillon-Aussel, 2018). The location of the priority development areas focused in this urban renaissance program are shown in Figure 6, together with another map showing the location of the 23 special wards corresponding to the former Tokyo municipality in the Eastern part of the Tokyo prefecture, for which the web site "skyscraperpage.com" is compiling information about high-rise buildings. Table III lists the number of buildings located in these 23 special wards, sorted according to their construction date in each of the 5 last decades since 1970, for three different height thresholds: 50, 100 and 150 m. It shows that there has been a major construction effort in the last three decades, with a peak in the period 2000-2009, especially for the very high-rise buildings (height above 100 and 150 m).

*Table III :Number of nigh-rise buildings built in the last five decades (from 1970 to 2019) in Tokyo (source: skyscraperpage.com)*

| Number of new skyscrapers | Decade | | | | |
|---|---|---|---|---|---|
| | 1970-1979 | 1980-1989 | 1990-1999 | 2000-2009 | 2010-2019 |
| h>150 m | 8 | 2 | 16 | 73 | 61 |
| h>100 m | 17 | 27 | 95 | 239 | 169 |
| h>50 m | 51 | 96 | 282 | 551 | 318 |

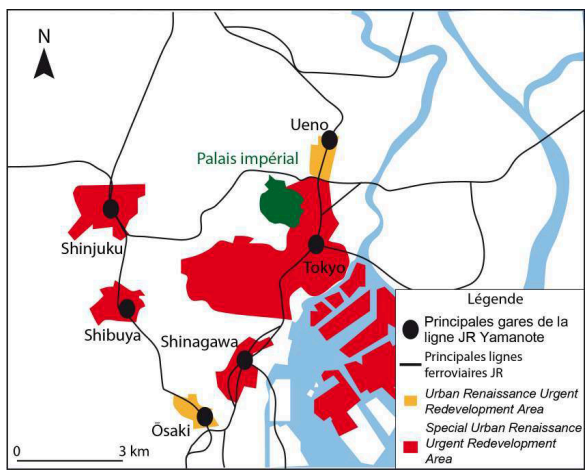
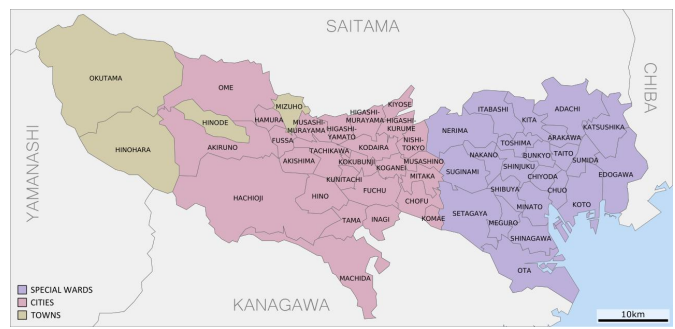

*Figure 6 Location of Priority Development Areas of the Urban Renaissance program (Left, from Languillon-Aussel, 2018), and of the 23 special wards (right part, light purple area, from fr.maps-tokyo.com) of the Tokyo prefecture within which skyscraper statistics are available*

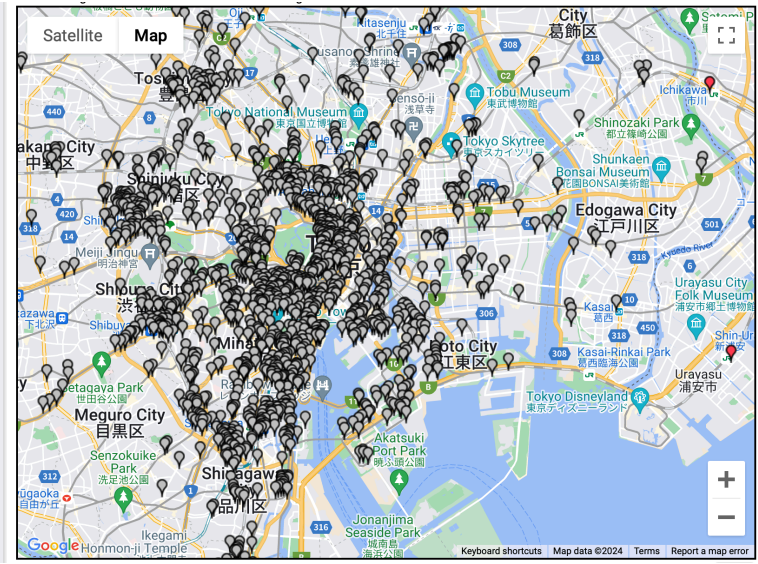
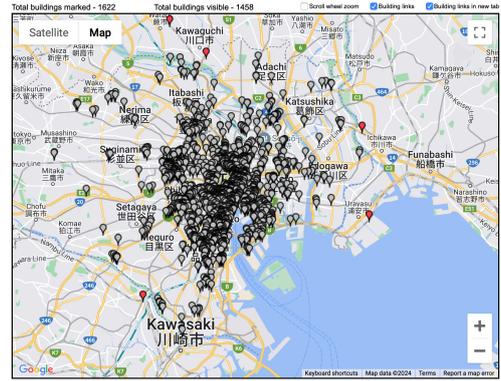

Buildings with h ≥ 50 m

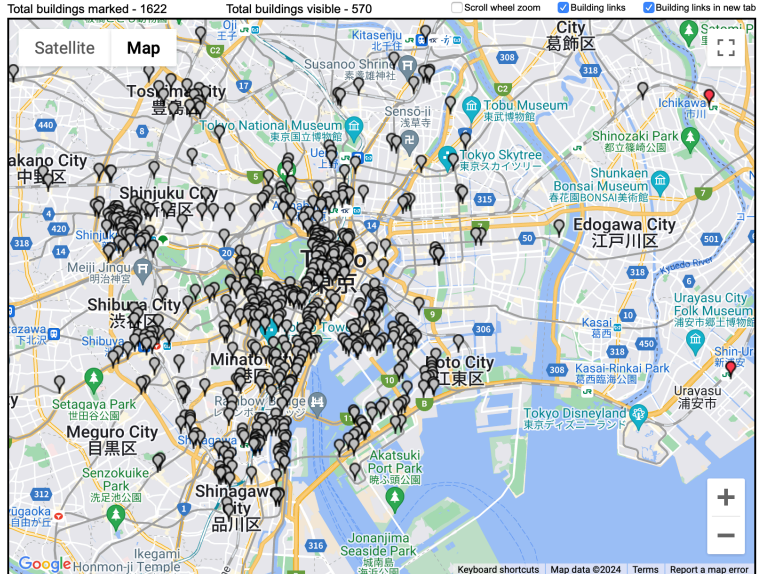
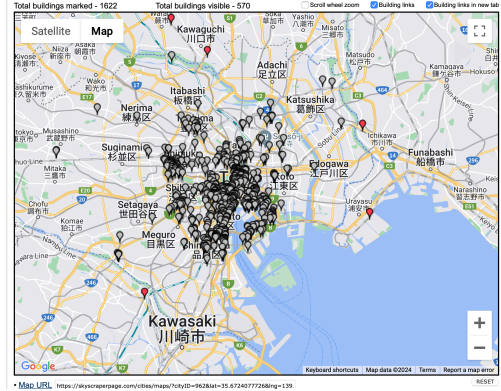

Buildings with h ≥ 100 m

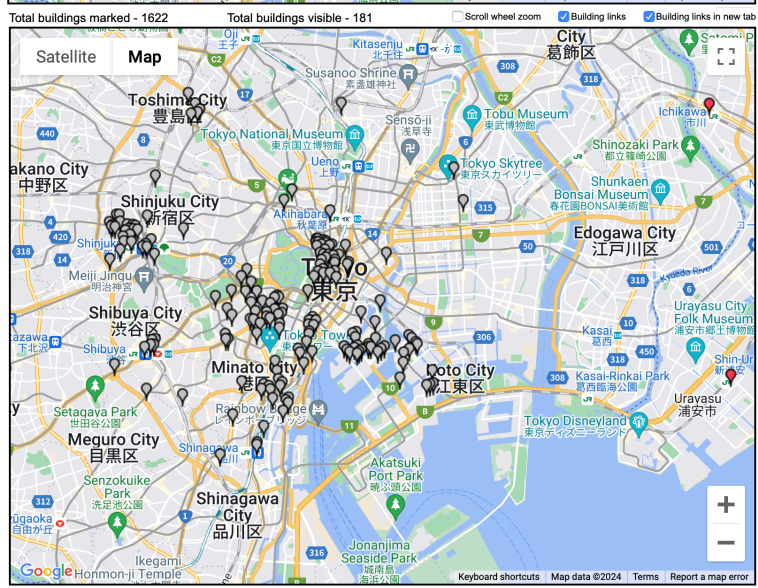
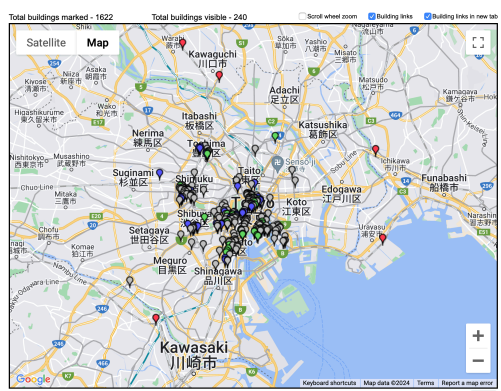

Buildings with h ≥ 150 m

*Figure 7: Location of high-rise buildings in the Tokyo central area, with three height thresholds (Top: 50 m, Middle: 100 m, Bottom: 150 m). The right part displays a wide zone, while the left maps zoom on the area with the largest number of high-rise buildings (all maps from skyscraperpage.com)*

The spatial location of all these buildings Is shown in Figure 7, with separate maps for the three different height thresholds. When crossing the two spatial and time information, it turns out that only a few special wards have been the place of major urban changes with a multiplication of high-rise buildings in priority redevelopment areas (see Figure 6). Furthermore, a comparison of the locations of sites with site response changes over the last two decades indicates an intriguing spatial correspondence with the location of new high-rise buildings. Sites with the largest reduction coincide with the location of new buildings with height larger than 100 to 150 m, while sites with noticeable and moderate reduction (yellow and orange pins in Figure 5) seem to better correspond to a larger number of intermediate height buildings (50 to 100 m), which are quite numerous also in the area comprised the two rivers (Sumida and Arakawa).

### 4.2  Possible links with site-city interaction mechanisms

This spatial and temporal coincidence could be at least partly explained by global structure-soil-structure interaction phenomena as described in the various studies shortly reported in the review section.

The overall reduction of site response over a broad frequency range could correspond to the overall decrease effect reported in numerical studies (from Clouteau & Aubry (2001) to Basnet et al., 2023). This kind of effect has been reported to be particularly efficient in case of coincidence between the site frequency and the building frequency (Bard et al., 2006; Boutin et al., 2015; Schwan et al., 2016). In the present case, the frequencies of high-rise buildings with height beyond 50 m are all below 1-2 Hz, and probably below 0.5 Hz for buildings exceeding 150 m. Meanwhile, as shown in Figure 2, the site peak frequency is generally between 0.5 and 1 Hz, and the site fundamental frequency is much lower, probably between 0.1 and 0.2 Hz. The dominant peak probably corresponds to the softer layers near the surface, while the fundamental one corresponds to the very thick sedimentary deposits, reaching several kilometres in the deepest part of the Kanto basin (see for instance Nimya et al, 2023). The stations with reduced site response are located within areas with many new buildings above 50 m height, and the associated reduction could thus be related with destructive interference effects between waves multiply scattered from the numerous high-rise buildings.

From another view point, as also illustrated in Figure 2, the site response within the Kanto basin is strongly affected by the presence of surface waves diffracted along the subsurface heterogeneities: the site term inferred from body waves only is significantly smaller than the site term derived from long windows including both body and surface waves). Several studies involving periodic or quasi-periodic "metamaterials" at the surface showed their major "blocking" effects for the propagation of Rayleigh waves in some frequency bands, clearly related to the natural frequencies of the surface oscillators. Since in the Kanto basin, long period surface waves are predominantly generated on basin edges several tens of kilometers apart from the central part of Tokyo (Kinoshita et al., 1992) part of the low-frequency reduction observed to the East and South-East of the Tokyo railway station might also be due to such a blocking effect of very high-rise buildings for long period surface waves generated on the western edge of the basin and traveling from West to East, leading to less local effects extending to the East of the re-cevelopped areas.

### 4.3  Other possible explanations

It is not the first time that site response changes are reported. Most often however, they are related either with sensor installation issues (changes in sub-sensor slab, ground level or buried at shallow depth, interaction with very near small structures, see Hollender et al., 2020), or with non-linear response issues. In the first case, they can be observed only at high frequency (i.e., beyond 10 Hz). In the second case, they are transient, and they most often recover after some lag time never exceeding several years (e.g., Richter et al., 2014; Gassenmeier et al., 2016; Viens et al., 2018; Bonilla et al., 2019). The considered area (Kanto) has indeed experienced a very large shock in the last two decades, but the observed changes (as illustrated in Figure 3c and d) do not exhibit neither a marked step after the great 2011 Tohoku earthquake, nor a slow recovery after some months/years. Moreover, similar changes are not observed in any station of the Chiba prefecture, which are closer to the source area and experienced larger motion, even though some of them are located on thick sedimentary deposits.

The consolidation of soft soils at rather shallow depth could contribute to the increase of soil stiffness, and therefore to the reduction of amplification. Such a consolidation could be due to water pumping, or to the construction of new buildings increasing the applied vertical stress. In both cases however, it is rather unlikely that such effects extend to large depths, i.e. exceeding 100 m or so, even for high-rise buildings or even skyscrapers. Meanwhile, the measured changes occur at low-frequency (0.2-1 Hz), corresponding to wavelengths in the range (200 – 1500 m) for a shear wave velocity of 200 m/s, or (500 – 2500 m) for a shear wave velocity of 500 m/s. The shear-wave velocity profile of the involved K-net stations is not available, but the J-SHIS model (SDLCM, 2019) and the Kanto basin model derived for multimodal inversion of surface waves (Nimya et al., 2023) both indicate surface wave velocities between 500 and 1000 m/s for the fundamental mode of Love waves in the frequency range [0.2, 0.8 Hz]: the depths implied by the steady velocity changes due to consolidation should therefore exceed 500 m, which seems quite unlikely.

Finally, another possibility could be the densification of underground structures (subway lines, deep foundations, etc.) which may be responsible by an overall near surface stiffness increase, which in turn could result in amplification decrease in relation to the increased impedance. This could in principle be investigated with a careful monitoring of the velocity changes at shallow depth (i.e., not exceeding 100 m).

## 5    Concluding comments

Given the topic of this WCEE2024 special session ("Early history of site effects evaluation and dynamic soil-structure interaction in earthquake engineering"), the aim of this contribution was to report how the unexpectedly large amplifications and durations observed in the Mexico City lake bed area triggered a series of numerical and experimental investigations on the feedback of building structures on urban, "free-field" ground motion, and to which extent the outcomes of such investigations are consistent with the analysis of site response changes in the central Tokyo area. The latter could be possible because of a) the installation, maintenance of K-net and KiK-net accelerometric networks since the mid-nineties (in addition to the JMA Shin-dokei stations), b) the sustained seismic activity level leading to an average number of around 130 recordings for a total of 106 stations with more than ten years operation in the four main prefectures of the Kanto area (Chiba, Kanagawa, Saitama, and Tokyo), c) the fully open availability of recordings, and d) the derivation of site-specific terms for each event through the generalized inversion scheme presented in Nakano et al. (2015).

In short, it is found that the sites located in or near the priority redevelopment areas of the "urban renaissance" program in Tokyo, where many new high-rise buildings have been erected since the turn of the century, exhibit a significant reduction of the low-frequency amplification (up to a factor of 3 reduction at some sites). Considering the correspondence between high-rise building frequencies (below 1 Hz) and site frequencies (fundamental mode below 0.2 Hz in relation with very thick – around 3 kilometres – sedimentary deposits, and largest amplification between 0.5 and 1 Hz in relation to softer soils at shallow depth), such a decrease is consistent with the outcomes of all kinds of investigations (numerical and experimental) carried on the last three decades about effects of multiple interaction between buildings and underground soil structure.

However, such a consistency, although intriguing, cannot be considered a definite proof that the observed reduction is actually due to "site-city interaction" effects. Several additional investigations should be performed to confirm (or eliminate) the SCI interpretation:

- Updating the site terms derived from GIT by considering the latest recordings since 2016: this would allow to have also long enough data from several JMA stations (see Figure 1) located in the area of interest, but installed too late to provide at least 10 years of continuous data in the latest generalized inversion performed by Nakano and colleagues
- Considering other types of site response estimates, such as SSR or HVSR, could be helpful also for analyzing the robustness of GIT inversion results. The first technique however requires the selection of an appropriate reference site, which is not easy for the Kanto plain, while the second is only a proxy for site amplification and might be biased by simultaneous SCI effects on horizontal and vertical components
- Considering more quantitative measurements of urban changes, involving for instance the percentage of built surface, the spatial distribution of building heights, volumes and mass (e.g., Yoshida and Omae, 2005; Jing et al., 2021)
- Investigating the near-surface (i.e., over the top few hundred meters) velocity changes implied by the new surface and underground constructions

Similar analysis should also be carried out in other areas which recently experienced some major urban developments, such as other cities in Japan (Osaka/kobe/Kyoto area, Nagoya, Sendai, …) or in other parts of the World (Mexico City, USA and Canada west coast, China, etc.), provided that within-cities seismological recordings are numerous enough over a long enough time period, and freely available as in Japan.